# Manipulation of magnetic skyrmions in continuous Ir/Co/Pt multilayers

M. Cubukcu[1,2*], S. Pöllath[3], S. Tacchi[4], A. Stacey[5], E. Darwin[5], C. W. F. Freeman[1,2], C. Barton[1], B.J. Hickey[5], C.H. Marrows[5], G. Carlotti[6], C.H. Back[7], and O. Kazakova[1]

[1]National Physical Laboratory, Teddington, TW11 0LW, United Kingdom

[2]London Centre for Nanotechnology, University College London, 17-19 Gordon Street, London, WC1H 0AH, United Kingdom

[3]Institut für Experimentelle Physik, Universität Regensburg, D-93040 Regensburg, Germany

[4]Istituto Officina dei Materiali del CNR (CNR-IOM), Sede Secondaria di Perugia, c/o Dipartimento di Fisica e Geologia, Università di Perugia, I-06123 Perugia, Italy

[5]School of Physics and Astronomy, University of Leeds, Leeds LS2 9JT, United Kingdom

[6]Dipartimento di Fisica e Geologia, Università di Perugia, Via Pascoli, I-06123, Perugia, Italy

[7]Technical University Munich, 85748 Garching, Germany

[*]murat.cubukcu@npl.co.uk

## Abstract

We show that magnetic skyrmions can be stabilized at room temperature in continuous $[Ir/Co/Pt]_5$ multilayers on $SiO_2$/Si substrate without prior application of electric current or magnetic field. While decreasing the Co thickness, tuning of the magnetic anisotropy gives rise to a transition from worm-like domain patterns to long and separate stripes. The skyrmions are clearly imaged in both states using Magnetic Force Microscopy. The density of skyrmions can be significantly enhanced after applying the "in-plane field procedure". In addition, we have investigated the phase diagram of a sample deposited in the same run, but onto a $SiN_x$ membrane using Lorentz transmission electron microscopy. Interestingly, this sample shows a different behaviour as function of magnetic field hinting to the influence of strain on the phase diagram of skyrmions in thin film multilayers. Our results provide means to manipulate skyrmion, further allowing for optimized engineering of skyrmion-based devices.

## Introduction

Low-dimensional topological spin textures in magnetic materials are technologically attractive since it is expected that they can be used in next generation storage devices as information carriers. One example is the magnetic skyrmion; a nanometre-sized topological defect representing a swirling spin texture. Their topological properties, efficient current-driving dynamics together with their nanoscale size and stable particle-like features make magnetic skyrmions promising candidates for carrying magnetic information in future high-density and low-power consumption spintronic devices [1-4].

Recently, heavy metal (HM) / ferromagnet (FM) multilayers hosting skyrmions deposited by magnetron sputtering have attracted attention. Indeed, the strong spin-orbit coupling (SOC) of the HM layer can lead to an antisymmetric exchange known as interfacial Dzyaloshinskii-Moriya interaction (iDMI) [5, 6], which plays a key role in the stabilization of magnetic skyrmions [7-10]. When iDMI is utilized for the formation of skyrmions, it is possible to control the nucleation processes and skyrmions properties using a variety approaches [11, 12]. In this way, relevant magnetic parameters, such as perpendicular magnetic anisotropy (PMA) or the DMI strength can be strongly modified to affect both their nucleation and properties (e.g. density, size, dynamics).

Although skyrmions have been observed at room temperature in HM/FM multilayers, in most cases their nucleation and stabilization require an injection current or external magnetic field [13-15]. Additionally, lithographically defined structures were used to confine single or multiple skyrmions depending on the geometry [16, 17]. Moreover, the investigations on how to enhance the skyrmion density are of great significance for achieving ultrahigh density spintronics devices [18-20].

In this work, we show that magnetic skyrmions can be stabilized at room temperature in continuous $[Ir/Co/Pt]_5$ multilayers on $SiO_2/Si$ substrate; external magnetic field, injection current or geometric confinement are not required to generate skyrmions. The magnetic, structural, and interfacial parameters of the multilayer are analysed using Vibrating sample magnetometer (VSM), X-ray reflectivity (XRR) and Brillouin light scattering (BLS). Imaging of skyrmions was performed by Magnetic force microscopy (MFM) and Lorentz transmission electron microscopy (LTEM). Tuning of the magnetic anisotropy by thinning the Co layer ($t_{Co}$) leads to a transition of the magnetic domain patterns from a worm-like state to long and separated stripes. The skyrmions are clearly observed in both states. In addition, we have also measured the phase diagram of a counterpart sample (deposited on a $SiN_x$ membrane) using LTEM and we find a clear difference for the identical multilayer stack deposited on different substrates which we tentatively attribute to the contribution of mechanical stress. We also report that the density of skyrmions can be significantly enhanced after undergoing an “in-plane field procedure”, where a high density of skyrmions can be detected after applying an in-plane magnetic field around 2 T and subsequently ramping it to zero. Magnetization curves showed the dependence of the perpendicular magnetic anisotropy (PMA) with the Co thickness, providing a direct link to understand the magnetic textures observed in the MFM images. These results could provide a criterion for designing the skyrmion magnetic thin films, which has a potential to advance the development of skyrmion-based magnetic devices.

## Sample fabrication and Characterization

The multilayers $[Ir(1.2nm)/Co(t_{Co})/Pt(1.3nm)]_{x5}$ ($t_{Co}$=0.4nm-0.8nm), were grown using DC magnetron sputtering in a high vacuum system. The samples grown on $SiO_2$/Si substrates were used to determine the magnetic, structural and interfacial properties using VSM, BLS and XRR as well as to image their magnetic textures using MFM. Additionally, the identical counterpart multilayers grown on $SiN_x$ membrane (deposited in the same run) were used for LTEM. The base pressure in the chamber before growth was of the order of $1x10^{-8}$ mbar, and a flow of 60 sccm/5.02 mTorr of Argon gas was used throughout the sputtering process. The different layers in the multilayer structure were grown in turn by moving the substrate over the top of the sputter guns for known periods of time, while applying a constant source current to the target materials. The target composition, gun position, source current, and subsequent power of the magnetron gun for each material is shown in Table S1 (see supplementary information), as well as the typical growth rates for each material. The separation between the sputter target and the sample substrate was a constant 7 cm during growth. The multilayer structure is schematically illustrated in Figure 1a. In the $[Ir/Co/Pt]_5$ multilayer film, interfacial DMI exists between $S_1$ and $S_2$ of two adjacent Co atoms close to heavy metals (Ir or Pt) with a strong SOC. The Hamiltonian can be expressed as $H_{DMI} = D_{12} \cdot (S_1 \times S_2)$ [21], where $D_{12}$ is the DMI vector as shown in Figure 1a. The sample structure was characterized using XRR (Fig.1b) and the resulting fringe pattern was simulated using GenX [22] confirming the multilayer structure. Fitting parameters are shown in S1 (see supplementary information).

To determine the strength of the iDMI we exploited BLS. BLS measurements from thermally excited spin waves (SWs) were performed in the backscattering geometry focusing about 150 mW of a monochromatic laser beam (wavelength λ = 532 nm) on the sample surface through a camera objective with numerical aperture NA = 0.24. The frequency of the scattered light was analysed by a Sandercock-type (3+3)-tandem Fabry-Perot interferometer. Due to the conservation of momentum in the light scattering process, the magnitude of the spin wave vector *k* is related to the incidence angle of light *θ*, by the relation *k* = 4π sin *θ*/λ. First, the dependence of the SW frequency as a function of the intensity of the in-plane applied field $\mu_0 H$ was measured at normal incidence, *i.e.* for k=0 rad/µm. (Fig.1c, dots). To quantitatively estimate the out-of-plane anisotropy constant $K_u$ and the gyromagnetic ratio γ, a best fit procedure of the experimental data (Fig.1c, red line) was performed using the Kittel equation:

$$\left(\frac{\omega}{\gamma}\right)^2 = \left[H \cdot \left(H - \frac{2}{M_S} K_u + 4\pi M_S\right)\right]$$

where the saturation magnetization was fixed to the value $M_S$=1.65·10$^3$ kA/m as measured by VSM. From this analysis the values $K_u$ =1.89·10$^6$ J/m$^3$ and γ= 176 GHz/T were obtained for $t_{Co}$=0.8 nm sample. The strength of the iDMI was quantitatively extracted by measuring the iDMI induced frequency asymmetry, $\Delta f$, for Damon-Eshbach (DE) modes propagating in opposite directions. BLS measurements were performed in the DE geometry applying an in-plane magnetic field $\mu_0 H$=1.5 T, sufficiently large to saturate the magnetization in the film plane and sweeping the in-plane transferred wave vector along the perpendicular direction. The top inset of Figure 1d shows the BLS spectra measured at *k*=2.25×10$^7$ rad/m. One can observe that the Stokes and anti-Stokes peaks are characterized by a sizeable frequency asymmetry, which reverses upon reversing the magnetic field direction. Figure 1d shows the SW frequency asymmetry, $\Delta f$, measured at *k*=1.67×10$^7$ rad/m and *k*=2.25×10$^7$ rad/m upon reversing the direction of the applied magnetic field, that is equivalent to the reversal of the propagation direction of the DE mode. The effective DMI constant, *D*, was determined by means of a linear fit (continuous red line) to the experimental data using the relation $\Delta f = \frac{2\gamma D}{\pi M_s} k$, fixing the gyromagnetic ratio and the saturation magnetization to the value obtained from the analysis of the BLS measurements as a function of $\mu_0 H$ and from VSM measurements, respectively. A value $D = (1.8 \pm 0.2) mJ/m^2$ was obtained, indicating that the right-handed chirality is favoured by the iDMI. In fact, for a Co/Pt stack where the Pt is the overlayer, we expect a right-handed chirality, in agreement with previous investigations [23].

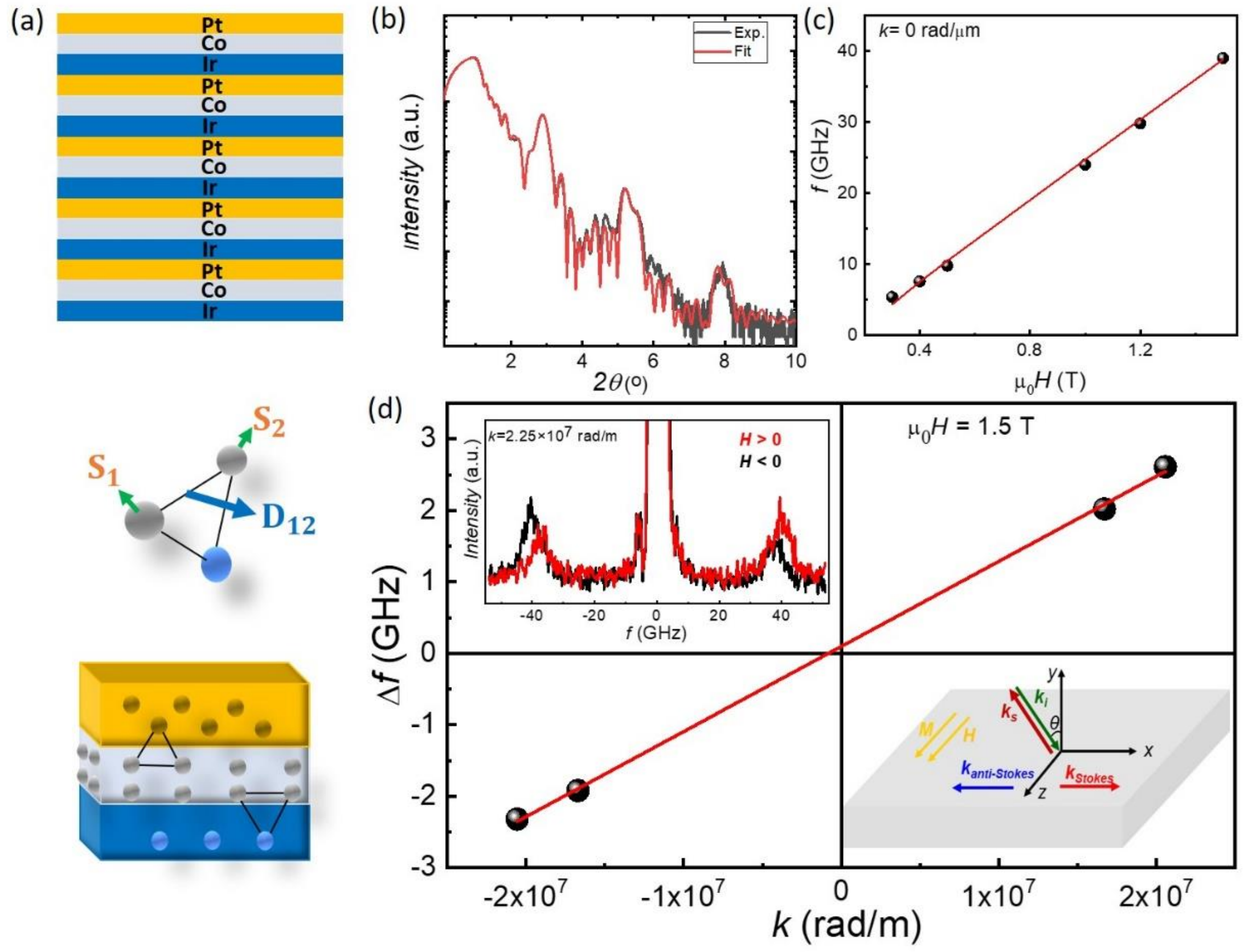

***Fig.1**.* **Structural and interfacial characterization of [Ir/Co/Pt]$_5$ multilayers for $t_{Co}$=0.8 nm at room temperature.** (a) Schematic diagrams of a multilayer film structure and the corresponding interfacial DMI produced between $S_1$ and $S_2$ of two adjacent Co atoms close to the Ir atoms or Pt atoms with a strong SOC. (b) XRR measurement results: intensity as a function of $2\theta^{\circ}$ incident angle. (c,d) BLS measurements. (c) Dependence of the spin waves (SW) frequency (*f*) as a function of the intensity of the in-plane applied field $\mu_0 H$ was measured at normal incidence, for *k*=0 rad/μm. (d) The SW frequency asymmetry, $\Delta f$, measured at *k*=1.67×10$^7$ rad/m and *k*=2.25×10$^7$ rad/m on reversing the direction of the applied magnetic field, that is equivalent to the reversal of the propagation direction of the DE mode. Top inset: BLS spectra measured at *k*=2.25×10$^7$ rad/m. Bottom inset: Schematic of BLS experiment. The sample is saturated in-plane by an external field $\mu_0 H$= 1.5 T, applied along the z-axis. Stokes and anti-Stokes events in the scattering process correspond to SW propagating with +k and -k, respectively.

## Results and Discussion

The configuration of the vertical magnetic texture has been investigated by MFM. MFM imaging of the multilayers was performed at room temperature with a NT-MDT Ntegra Aura scanning probe microscope (SPM) [24-26]. The system is fitted with an electromagnet, which allows applying out-of-plane magnetic fields up to 115 mT during scanning. Low moment tips (NT-MDT MFM-LM) were chosen in order to minimise probe–sample interaction. To image the magnetic domains patterns without any prior applied magnetic field, the samples were imaged in the as-grown state for $t_{Co}$=0.8 nm sample (Fig.2a). The MFM images show that the magnetization is broken up into small domains of a worm-like configuration. Some skyrmions are also clearly observed among the worm-like textures, see the indications by the dashed black arrows in Fig.2a. As the measurements were performed before cycling the magnetic field, these images reveal that no prior stabilizing magnetic field or injection current are required to generate skyrmions. Therefore, skyrmions at zero field can be spontaneously stable even for samples in the as-grown state. Then, in order to explore the different processes that can stabilize skyrmions or/and manipulate density of skyrmions, the sample was imaged after applying an in-plane magnetic field around 2 T and subsequently turning off the in-plane magnetic field (Fig.2b.). In the following, we refer to this sequence as the “in-plane field procedure”. In fact, in previous investigation, it has been reported that the applied in-plane component of the magnetic field will affect the concentration of skyrmions [27]. In Fig.2b, we show the MFM images at zero field after application of the “in-plane field procedure”. This procedure is highly favourable for skyrmion formation and increases their density synergistically, toggling a maximal skyrmions area value of ≈0.35 $\mu m^2$. Next, the MFM images were obtained under different applied out-of-plane magnetic field $\mu_0 H$; an example of the images at $\mu_0 H = 32\ mT$ shown in Fig. 2c. In Fig.2d, we show that the skyrmions' area decreases linearly with the applied $\mu_0 H$. The circularity is almost constant at low field and increases slightly when the $\mu_0 H$ is increased.

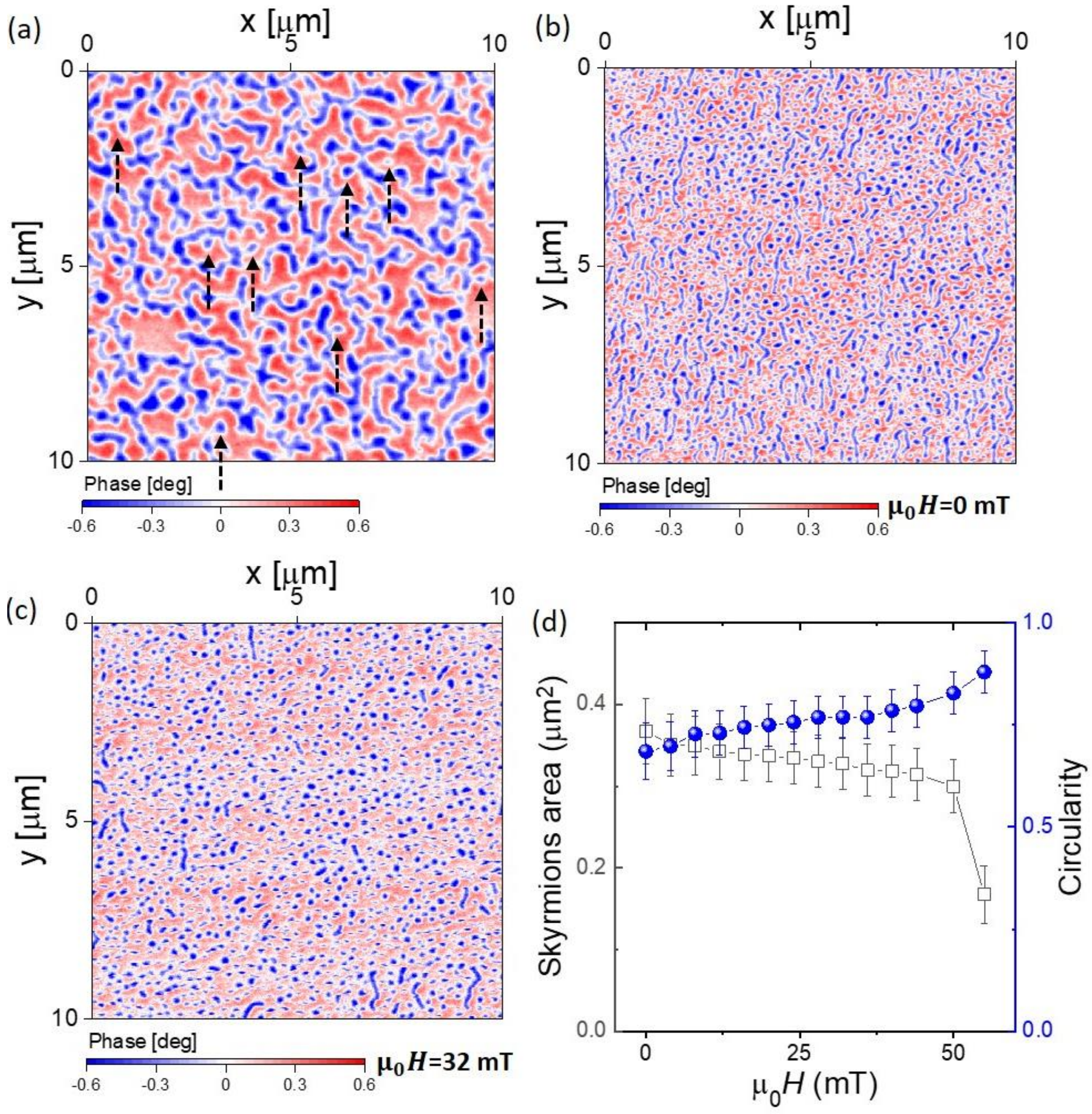

***Fig.2***. **MFM measurements on [Ir/Co/Pt]5 multilayers for $t_{Co}$=0.8nm at room temperature.** (a) The MFM images were acquired in the as-grown state. Red and blue contrast represent the out-of-plane magnetizations of opposite direction. Some skyrmions are indicated by dashed black arrows. (b) As grown-state modified by the "in-plane field procedure". (c) An example of the evolution of skyrmions vs the perpendicular applied magnetic field for $\mu_0 H = 32mT$. (h) A plot showing the area of the skyrmions (open square) and circularity (blue spheres) vs $\mu_0 H$.

We further study the effect of Co thickness on the magnetic properties of [Ir/Co($t_{Co}$)/Pt]5 multilayers (Fig. 3). For $t_{Co}$=0.6 nm, the magnetic domains exhibit a clear worm-like configuration, though some individual skyrmions can be seen in the as grown-state (Fig. 3a). By further reducing the Co layer thickness ($t_{Co}$=0.4 nm), we observe a transition from the worm-like to long stripes magnetic domain pattern (Fig. 3b). By reducing the Co thickness to 0.4 nm, we also observe a smaller size of skyrmions in the as-grown state (Fig. 3b). In addition, in Fig. 3c, we show the skyrmions' area versus $t_{Co}$ at zero field after the “in-plane field procedure”. The skyrmions' area decreases whith decreasing $t_{Co}$ and the circularity remains almost constant. To understand the thickness dependent properties, we refer to the magnetisation measurements. The out-of-plane and in-plane magnetization curves (normalized to the saturation magnetization $M_S$) are summarized in Fig. 3d. For thicker Co samples (0.8 nm and 0.6 nm), the out-of-plane hysteresis shows a tail feature, whilst thinner sample (0.4 nm) present a more square-shaped loop. The anisotropy field (see dashed arrows), which is obtained from the in-plane magnetization curve at saturation is higher for the samples with thinner Co layers, indicating an increased PMA.

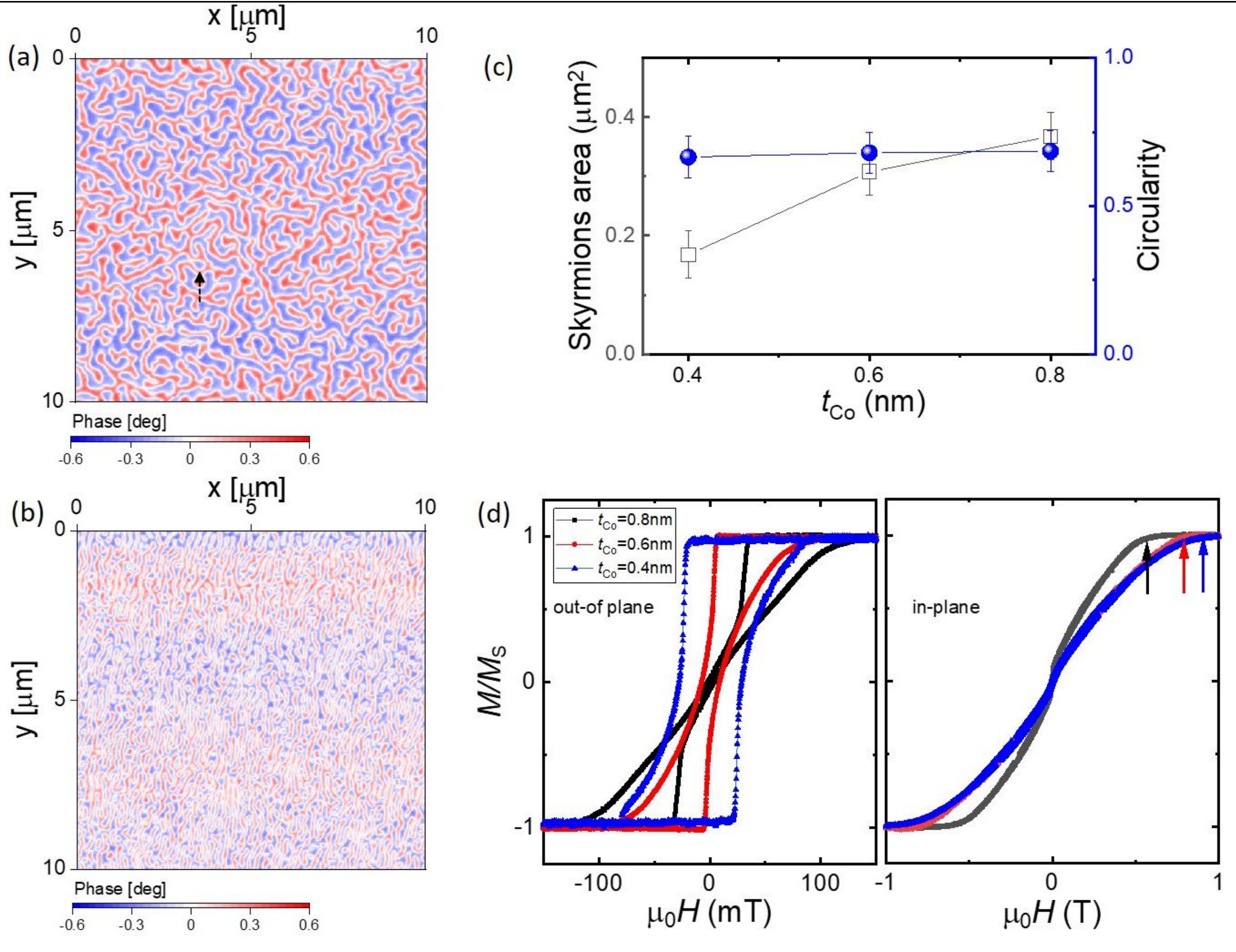

***Fig.3.*** **MFM and magnetization measurements vs Co thickness ($t_{Co}$) at room temperature.** The MFM images were acquired in the as-grown state for $t_{Co}$=0.6nm (a) and $t_{Co}$=0.4nm (b). (c) A plot showing the area of the skyrmions and the circularity vs $t_{Co}$ at zero field after the “in-plane field procedure”. (d) Normalised hysteresis curves, $M/M_S$ vs the external magnetic field $\mu_0 H$, in both the out-of plane (left) and in-plane (right) directions for $t_{Co}$=0.8 nm, $t_{Co}$=0.6 nm and $t_{Co}$=0.4 nm.

As a final step in our investigation, using Lorentz transmission electron microscopy (LTEM, Tecnai-F30) we have examined a $[Ir/Co/Pt]_5$ sample with Co layer thickness $t_{Co}$=0.8 nm deposited onto an electron beam transparent $SiN_x$ membrane under the same experimental condition (in the same run) as the previous samples. Since we investigate Néel type skyrmions, the sample has been tilted by 20° in all images. The temperature was set in a liquid nitrogen sample holder and controlled by using an additionally installed heater. The multilayer has been investigated in a temperature regime of approximately from -175 to 75 °C. At all temperatures, the externally applied magnetic field $\mu_0 H$ has been ramped from zero to 180 mT (after zero-field cooling). Figure 4a shows the phase diagram resulting from a zero-field cooling procedure. Here, a designated temperature is set and reached in zero magnetic field. Subsequently, the magnetic field is ramped towards high fields and images are recorded as indicated by the black dots in Fig. 4a. After the measurements, the sample is warmed up again and a new target temperature is set. In our LTEM images, we observe worm-like domains in most regions of the phase diagram below the critical field, where the sample becomes entirely field polarized. We denote this region 'the helical phase' (Hel) in reminiscence of the order helical phase in B20 bulk materials [28]. This region is coloured green in Fig. 4a. Upon increasing the magnetic field, the domain state becomes scarcer and is mostly replaced by the field polarized state (light green). Only in a very narrow region of magnetic field and temperature space (light blue), some (very few) skyrmions are observed in this sample, see red box in Fig. 4b. Sample images from the different regions are shown in Fig.4b-d: (b) mixed field polarized-skyrmion state with some residual worm domains, (c) worm domains and field polarized state, (d) field polarized state, (e) worm domain state in close to zero magnetic field. The results obtained by LTEM are distinctly different from the MFM results and can be traced back to the use of a different substrate for the LTEM experiments. It is known that metallic films deposited onto thin low stress SiNx substrates lead to a warping of the substrate that can significantly alter the magnetic state [29] through stress induced anisotropy changes.

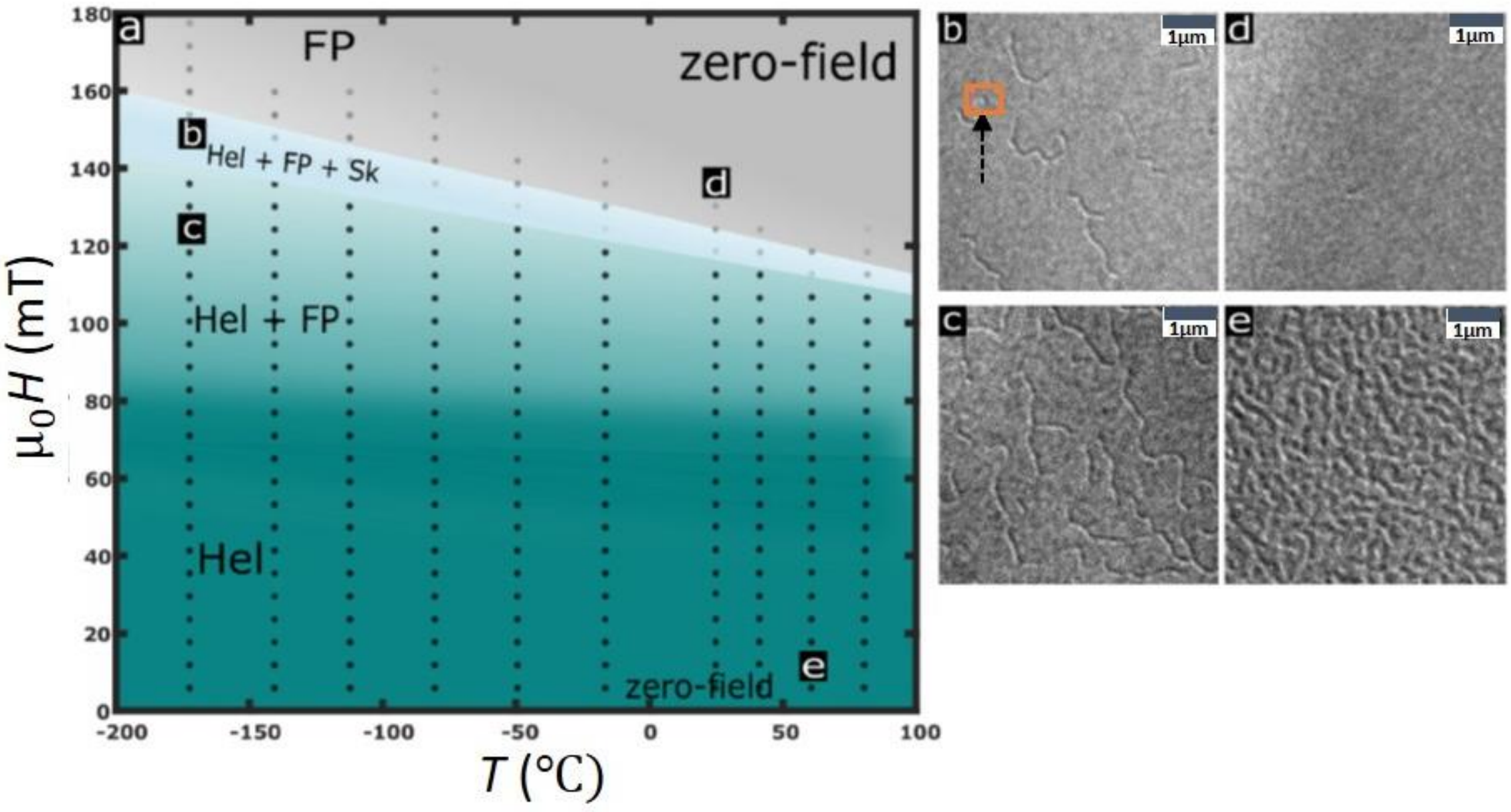

***Fig.4.*** **LTEM study of [Ir/Co/Pt]$_5$ multilayers for $t_{Co}$=0.8 nm.** (a) The phase diagram resulting from a zero-field cooling procedure in the as-grown state. (b-e) Sample images from the different regions as indicated in Fig. 4a: mixed field polarized-skyrmion state (FP+Sk) with some residual worm-domains (helical phase (Hel)) (b), worm-domains and field polarized state (FP) (c), FP state (d), worm-domain (Hel) state in close to zero magnetic field (e). A skyrmion indicated by dashed black arrow.

## Conclusion

In summary, we investigated the formation of magnetic domains in $[Ir/Co/Pt]_5$ multilayers using MFM and LTEM images. The magnetic skyrmions can be stabilized at room temperature without prior application of either electric current nor magnetic field. By thinning the Co thicknesses, we observed a transition from worm-like magnetic domains pattern to long stripes. The skyrmions are also clearly observed in both states. Significantly, the high density of skyrmions are imaged after undergoing the “in-plane field procedure”. Our results could provide a criterion for designing the skyrmion magnetic thin film, which may advance the development of skyrmion-based magnetic devices.

## Acknowledgements

The project 17FUN08 TOPS has received funding from the EMPIR programme co-financed by the Participating States and from the European Union's Horizon 2020 research and innovation programme. This project was also supported by the UK government department for Business, Energy and Industrial Strategy through NMS funding (Low Loss Electronics) and the UK national Quantum Technologies programme. C.W.F.F. thanks EPSRC for support through EPSRC DTP Case studentship (EP/T517793/1).

## References

[1] J. Sampaio, V. Cros, S. Rohart, A. Thiaville, A. Fert, 8, 839−844 (2013).

[2] A. Fert, N. Reyren, V. Cros, Nat. Rev. Mater. **2**, 17031, (2017).

[3] M. Cubukcu et al. Appl. Phys. Lett., **112**, 262409 (2018).

[4] J. Grollier, D. Querlioz & M.D. Stiles, Proc. IEEE, **104**, 2024-2039 (2016).

[5] I.A. Dzyaloshinsky, J. Phys. Chem. Solids **4**, 241–255 (1958).

[6] T. Moriya, Phys. Rev. **120**, 91–98 (1960).

[7] A. Thiaville et al. Europhys. Lett. **100**, 57002 (2012).

[8] R. Wiesendanger, Nature Rev. Mat. **1**,16044 (2016).

[9] Y. Yoshimura et al. Nat. Phys. **12**, 161 (2016).

[10] C. Moreau-Luchaire et al. Nat. Nanotechnol. **11**, 444-448 (2016).

[11] M. Cubukcu et al. Phys. Rev. B., **93**, 020401 (2016)

[12] A. Sud et al. https://arxiv.org/abs/2105.03976 (2021).

[13] G. Chen et al. Appl. Phys. Lett. **102**, 222405 (2013).

[14] A. Hrabec et al. Nat. Commun. **8**, 15765 (2017).

[15] S. Woo et al. Nat. Mater. **15**, 501–506 (2016).

[16] O. Boulle et al. Nat. Nanotechnol. **11**, 449–454 (2016).

[17] P. Ho et al. arXiv:1709.04878 (2017).

[18] A. Soumyanarayanan et al. Nat. Mater.**16**, 898−904 (2017).

[19] L. Wang et al. ACS Appl. Mater. Interfaces **11**, 12098−12104 (2019).

[20] X. Wang et al. Phys. Rev. **104**, 064421 (2021).

[21] A. Fert and P.M. Levy, Phys. Rev. Lett. **44**, 1538 (1980).

[22] M. Björck and G. Andersson, J. Appl. Cryst. **40**, 1174-1178 (2007)

[23] M. Kuepferling et al. arXiv:2009.11830v2 (2021)

[24] O. Kazakova et al. J. Appl. Phys. **125**, 060901 (2019)

[25] A. F. Scarioni et al. Phys. Rev. Lett. **126**, 077202 (2021)

[26] H. Corte-León et al. Small **16** (11), 1906144 (2020)

[27] S. Zhang, J. Zhang, Y. Wen, E. M. Chudnovsky, and X. Zhang, Commun. Phys. **1**, 36 (2018).

[28] A. Fert, V.Cros, J. Sampaio, Nat. Nanotechnology, **8**, 152-156 (2013).

[29] C. Dietrich et al. Phys. Rev. B **77**, 174427 (2008)

## Supplementary information

### S1. Growth conditions

The sample structure was characterized using X-ray reflectometry (XRR) (Fig.1b in the main manuscript) and the resulting fringe pattern was simulated using GenX[1] confirming the multilayer structure is as follows: Ta(18.75)/$Ta_2O_5$(6±5)/Pt(100±2)/Co(8±2)|Pt(13±2)/[Ir(11.77±0.6)/Co(8.8±0.7)/Pt(13.2 ±0.9)]$_{x5}$/Pt(32.9±0.3) where all thicknesses are provided in angstroms, Å, and the error on the Ta thickness found from the GenX fit is negligible[2]. The XRR fringes are shown in Fig.1b in the main manuscript, and with the GenX fit included in Fig.1b.

| Target Material | Magnetron Gun Position | Source Current (mA) | Power (W) | Typical Growth Rate (Å/s) |
|---|---|---|---|---|
| Tantalum (Ta) | 8 | 50 | 14 | 1.15 |
| Platinum (Pt) | 4 | 25 | 9 | 1.64 |
| Cobalt (Co) | 2 | 50 | 15 | 0.70 |
| Iridium (Ir) | 6 | 25 | 9 | 0.91 |

***Table. S1.*** The target materials and sputtering parameters used to grow the magnetic multilayers.

[1]GenX - https://aglavic.github.io/genx/

[2] ± 2 x$10^{-6}$ Å.

**S2. LTEM field-polarized cooling measurements**

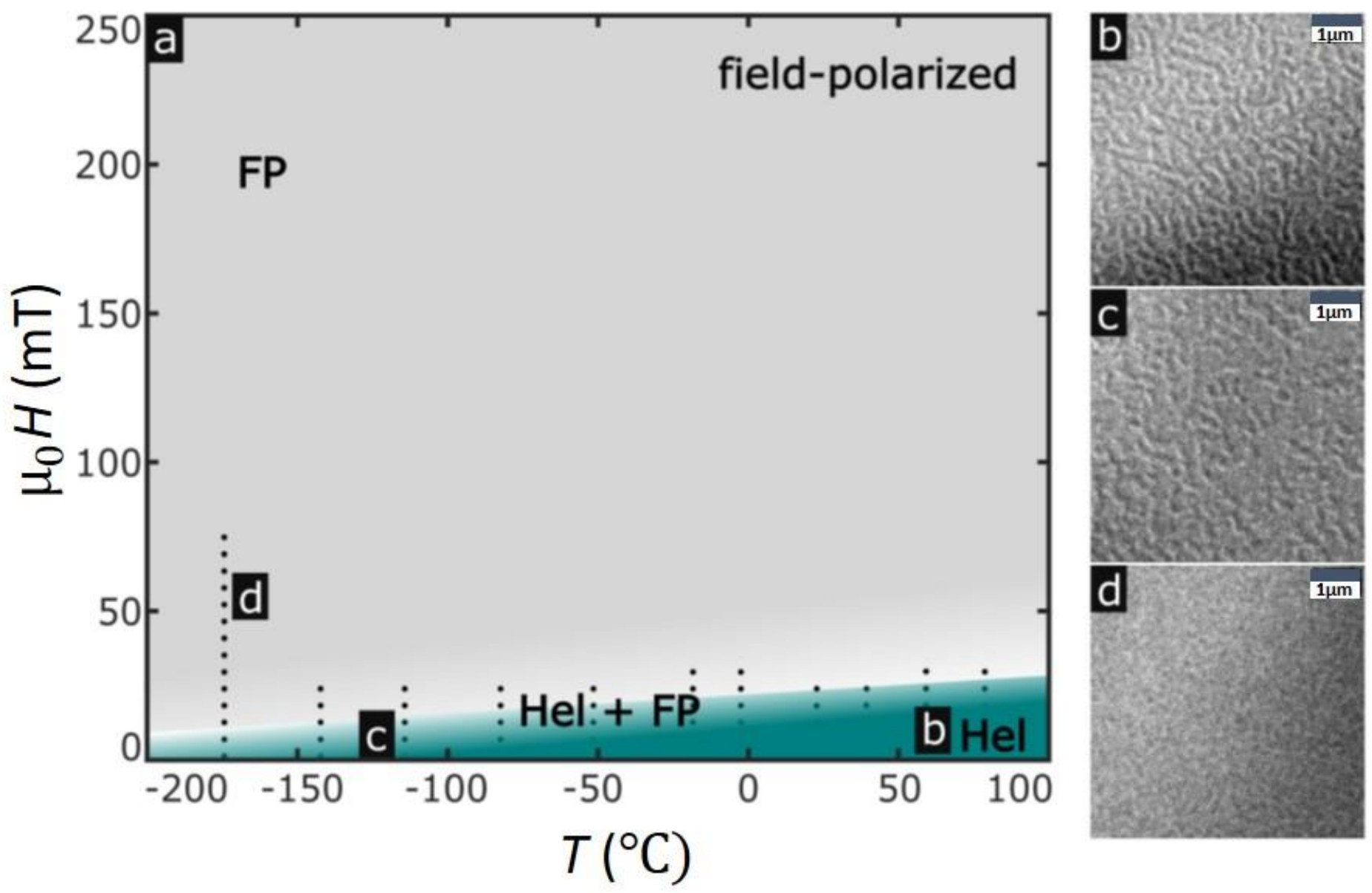

***Fig.S2.*** **(a) The phase diagram resulting from a field-polarized (FP) cooling procedure on multilayers for $t_{Co}$=0.8 nm.** Here, we field cool in a magnetic field of 250 mT and the reduce the field to zero. (b-d) The image series is taken from the different regions as indicated in Fig. S2a, show the hysteretic behaviour of the sample. In essence, we observe the field polarized (FP) state in most regions (d) in the magnetic field / temperature phase space and only close to zero magnetic field worm-like domains (b-c) are recovered.